\documentclass[sigconf]{acmart}

\usepackage{balance}
\usepackage{booktabs} \usepackage{enumitem}
\usepackage{listings}
\usepackage{tabularx}
\usepackage{xspace}
\usepackage[T1]{fontenc}
\usepackage[utf8]{inputenc}
\PassOptionsToPackage{hyphens}{url}

\renewcommand\footnotetextcopyrightpermission[1]{} 

\def\shownames{1}

\usepackage{pifont}\newcommand{\cmark}{\ding{51}}\newcommand{\xmark}{\ding{55}}\newcommand{\ccmarkl}{\multicolumn{1}{|c|}{\cmark}}\newcommand{\ccmarkm}{\multicolumn{1}{c|}{\cmark}}\newcommand{\cxmarkl}{\multicolumn{1}{|c|}{\xmark}}\newcommand{\cxmarkm}{\multicolumn{1}{c|}{\xmark}}
\newcommand{\dOne}{\ding{182}}
\newcommand{\dTwo}{\ding{183}}
\newcommand{\dThree}{\ding{184}}
\newcommand{\dFour}{\ding{185}}

\newcommand{\SafeKeeper}{\textsf{{\small{SafeKeeper}}}\xspace}
\newcommand{\SafeKeeperCaption}{\textsf{SafeKeeper}\xspace}

\newcommand{\SafeKey}{SafeKey\xspace}

\setlist[itemize]{leftmargin=*}

\begin{document}
\title[SafeKeeper: Protecting Web Passwords Using TEEs]{SafeKeeper: Protecting Web Passwords \\using Trusted Execution Environments}

\ifdefined\shownames

\author{Klaudia Krawiecka}
\affiliation{  \institution{Aalto University, Finland}
}
\email{kkrawiecka@acm.org}

\author{Arseny Kurnikov}
\affiliation{\institution{Aalto University, Finland}
}
\email{arseny.kurnikov@aalto.fi}

\author{Andrew Paverd}
\affiliation{\institution{Aalto University, Finland}
}
\email{andrew.paverd@ieee.org}

\author{Mohammad Mannan}
\affiliation{\institution{Concordia University, Canada}
}
\email{m.mannan@concordia.ca}

\author{N. Asokan}
\affiliation{\institution{Aalto University, Finland}
}
\email{asokan@acm.org}

\renewcommand{\shortauthors}{K. Krawiecka et al.}

\fi

\begin{abstract}
  Passwords are undoubtedly the most dominant user authentication
  mechanism on the web today. Although they are inexpensive and
  easy-to-use, security concerns of password-based authentication
  are serious. Phishing and theft of password databases are two
  critical concerns. The tendency of users to re-use
  passwords across different services exacerbates the impact of these
  two concerns. Current solutions addressing these concerns are not
  fully satisfactory: they typically address only one of the two concerns;
  they do not protect passwords from \emph{rogue servers}; they do not provide any \emph{verifiable evidence} of their (server-side) adoption to users; and
  they face \emph{deployability} challenges in terms of the cost for
  service providers and/or ease-of-use for end users.

  We present \SafeKeeper, a comprehensive approach to
  protect the confidentiality of passwords in web authentication systems. Unlike previous
  approaches, \SafeKeeper protects user passwords against very
  strong adversaries, including rogue servers and sophisticated external
  phishers. It is relatively inexpensive to deploy as it (i) uses widely
  available hardware security mechanisms like Intel SGX, (ii) is integrated
  into popular web platforms like WordPress, and (iii) has small
  performance overhead. We describe a variety of challenges in
  designing and implementing such a system, and how we overcome them. Through an
  86-participant user study, and systematic analysis and experiments, we
  demonstrate the usability, security and deployability of
  \SafeKeeper, which is available as open-source.

\end{abstract}

\maketitle

\section{Introduction}
\label{sec:introduction}

Passwords are by far the most widely used primary authentication mechanism on the web. 
Although many alternative schemes have been proposed, none has yet challenged the dominance of passwords.
In the evaluation framework of authentication mechanisms by Bonneau et al.~\cite{Bonneau2012a}, passwords have the best overall \emph{usability}, since they are easy to understand, efficient to use, and don't require the user to carry additional devices/tokens.
They also excel in terms of \emph{deployability}, since they are compatible with virtually all servers and web browsers, incur minimal cost per user, and are accessible, mature, and non-proprietary.
However, in terms of \emph{security}, passwords are currently a comparatively poor choice.
The two most critical security concerns, leading to the compromise of a large number of passwords, include: (i) phishing of passwords from users, and (ii) password database breaches from servers.  

Phishing attacks are prevalent in the wild, affecting home and business users alike (cf.\ APWG~\cite{apwg-phishing}, PhishTank~\cite{phishtank-stats}), and increasingly using TLS certificates from browser-trusted CAs (see e.g.~\cite{netcraft-letsenc-comodo}). 
While advanced anti-phishing solutions exist~\cite{Cui2017,Marchal2016}, and will improve over time, they alone cannot address the password confidentiality problem adequately, because users may unknowingly or inadvertently disclose passwords to malicious servers, which can then use the collected passwords at other servers. 
Note that an estimated 43--51\% of users reuse passwords across different services~\cite{Das2014} (see also more recent reuse analyses~\cite{Jaeger2016, Han2016, Wang2017}).

Password database breaches are increasingly frequent: hundreds of millions of passwords have been leaked in recent years (see e.g.~\cite{haveibeenpwned, pwd-vigilante}). 
Users are completely powerless against such breaches.
Password breaches are commonly dealt with by asking users to quickly reset their passwords, which is not very effective~\cite{Huh2017}.
Expecting users to select a different strong password for every site is unrealistic~\cite{Mickens2014}, and thus widespread password reuse makes these breaches problematic beyond the sites where the actual leak occurred~\cite{Jaeger2016}.
Stolen passwords can also lead to large-scale breaches of privacy/security sensitive data~\cite{verizon-analysis}.
Several recent solutions (e.g.~\cite{BirrPixton2016, Brekalo2016, Krawiecka2016, Everspaugh2015})
have been proposed to address password database breaches. 
Some of them (e.g.~\cite{BirrPixton2016, Brekalo2016, Krawiecka2016}) make use of hardware-based trusted execution environments (TEEs) on the server side, but none
can protect password confidentiality against \emph{rogue} servers (i.e.\ compromised servers, or servers belonging to malicious operators).

Designing effective solutions to protect passwords against rogue servers poses multiple technical challenges in terms of security (How to hide passwords from the authenticating server itself? How to rate-limit password testing by the server?);
usability (How to minimize the burden on users? How to support login from diverse user devices?);
user-verifiability (How to notify users when the solution is active?);
performance (How to realize this at scale?);
and deployability (How to allow easy/inexpensive integration with popular website frameworks?).
Bonneau et al.~\cite{Bonneau2012a} observed: ``many academic proposals have failed to gain traction because researchers rarely consider a sufficiently wide range of real-world constraints''.

We present \SafeKeeper, a comprehensive system for protecting password \emph{confidentiality} in web services.
Unlike all previous proposals, \SafeKeeper defends against both phishing and password database theft, even in the case of rogue servers.
At its core, \SafeKeeper relies on a \emph{server-side password protection service} that computes a keyed one-way function on the passwords before they are stored in the database.
This service is protected by a server-side TEE, which isolates the service from all other software on the server.  
\SafeKeeper includes a novel rate-limiting mechanism to throttle online guessing attacks at a rogue server.

\SafeKeeper's client-side functionality, which is implemented as a web browser addon, enables end users to detect whether a web server is running the \SafeKeeper password protection service within a server-side TEE, and to establish a secure channel from their browsers directly to it.
This mechanism \emph{assures} a user that it is safe to enter the password as it will be accessible only to the \SafeKeeper password protection service on the server. 
Unlike other client-side assurance approaches that rely on reliably identifying servers (e.g.\ by checking URLs or TLS certificates), or verifying their expected functionality, our approach relies only on signaling users via the \SafeKeeper browser addon, and thereby allowing them to correctly identify the server-side code that will process the password.
As long as users correctly recognize \SafeKeeper's client-side signaling, and enter their passwords only to \SafeKeeper-enabled web servers, confidentiality of their password is guaranteed,  \emph{even if users mistook the identity of the server (e.g.\ phishing), or the server is malicious/compromised (rogue server)}.
As such, \SafeKeeper may present a significant shift in phishing avoidance and password protection.

Our design considers \emph{deployability} as a primary objective.
We demonstrate this by developing a fully-functional implementation of \SafeKeeper using Intel's recent Software Guard Extensions (SGX), and integrating this with minimal software changes into \texttt{PHPass}, the password hashing framework used in popular platforms like WordPress, Joomla, and Drupal, which as of April 2018 accounted for over 35\% of the Alexa top 10-million websites~\cite{w3techs}.  
\SafeKeeper's client-end functionality does not depend on any additional device/service/hardware feature, and thus can be implemented in most user devices/OSes, including smartphones.
Our implementation is available as open-source software~\cite{SafeKeeper-GitHub}.

Our contribution is \SafeKeeper, including its:
\begin{itemize}
\itemsep0em
\item \textbf{Design}: As a password-protection scheme featuring
\begin{itemize}
\itemsep0em
\item a \textbf{server-side password protection service} using off-the-shelf trusted hardware to protect users' passwords, \emph{even against rogue servers} (Sections~\ref{sec:design-tee}--\ref{sec:design-attestation}), and
\item a novel \textbf{client-side assurance mechanism} allowing users to easily determine if it is safe to enter passwords on a web page (Section~\ref{sec:design-ui}). Our mechanism relies only on verifying whether the server runs \SafeKeeper \emph{without having to verify the server's identity or correct behavior}.
\end{itemize}

\item \textbf{Implementation and integration:} A full open-source implementation of (i) server-side functionality using Intel SGX, and integration into \texttt{PHPass} to support several popular web platforms, and 
(ii) client-side functionality realized as a Google Chrome browser addon (Section~\ref{sec:implementation}).

\item \textbf{Analysis and evaluation:} A comprehensive analysis of security guarantees (Section~\ref{sec:evaluation-security}), an experimental evaluation of performance (Section~\ref{sec:evaluation-performance}), deployability in real-world platforms (Section~\ref{sec:evaluation-deployability}), and validation of effectiveness of the client-side assurance mechanism via an \emph{86-participant user study} (Section~\ref{sec:evaluation-usability}). 

\item \textbf{Extensions and variations:} Several extensions and enhancements for providing backup and recovery mechanisms, scaling to multiple servers, and protecting email addresses (and potentially most other privacy-sensitive identity tokens) in addition to passwords (Section~\ref{sec:extensions}).

\end{itemize}

\section{Preliminaries}
\label{sec:preliminaries}

\subsection{Storing Passwords}
\label{sec:preliminaries-passwords}

The current best-practice for storing passwords is for the server to compute a \emph{one-way} function on the password (e.g.\ a cryptographic hash), and store only the result in the database.
When the user logs in, the supplied password is passed through the same one-way function and compared to the value in the database.
Although an adversary who obtains the database cannot reverse the one-way function, he can guess candidate passwords and apply the same one-way function in order to test his guesses.
Since passwords are not strong secrets (e.g.\ compared to cryptographic keys), this type of brute force guessing attack is often feasible.
The adversary can even speed up this attack using \emph{rainbow tables} -- pre-computed tables of hashed passwords.
If multiple users choose the same password, the results of the one-way function would also be the same.

To prevent the use of rainbow tables and avoid revealing duplicate passwords, it is customary to use a \emph{salt} -- a random number unique to each user that is concatenated with the password before being hashed.
However, since salt values are stored in the database, an adversary who obtains this database can still mount brute-force password guessing attacks against specific users by concatenating his guesses with the corresponding salt values.

Another layer of defence is to also use a \emph{pepper}~\cite{Blocki2016} -- a random value included in every password hash but not stored in the password database.
The pepper is usually the same for each user e.g.\ Dropbox uses a global pepper to protect their users' passwords~\cite{dropbox-passwords}.
Without knowing the pepper value, the adversary cannot test password guesses against a stolen database.
Therefore, the challenge is to protect the pepper from an adversary (who may compromise the server) whilst still being able to use it when users log in.
For example, Dropbox has indicated~\cite{dropbox-passwords} that in future they may store the pepper in a hardware security module (HSM).

\subsection{Trusted Execution Environments (TEEs)}
\label{sec:preliminaries-tee}

A TEE provides isolated execution for security-sensitive code.
Code running within the TEE (Trusted Applications, TAs) has exclusive access to data held within the TEE.
A recent example is Intel's Software Guard Extensions (SGX).

\subsubsection{Intel Software Guard Extensions}
\label{sec:preliminaries-sgx}

SGX is a set of CPU extensions available on the Skylake processor microarchitecture and onwards.\footnote{\url{https://software.intel.com/en-us/sgx}}
The new SGX instructions allow a userspace application to establish a hardware-enforced TEE, called an \emph{enclave}.
The enclave runs in the application's virtual address space, but after it has been initialized, only the code inside the enclave is allowed to access enclave data.
The application can call enclave functions (called \texttt{ecalls}) via well-defined entry points.
Enclave data is stored in a special region of memory, called the Enclave Page Cache (EPC), which can only be accessed by the CPU.
When any enclave data leaves the CPU (e.g.\ is written to DRAM), it is encrypted and integrity-protected, using a key accessible only to the CPU~\cite{Gueron2016}.
The enclave's data is therefore protected against privileged software (including the OS/hypervisor), and hardware attacks (e.g.\ snooping on the memory bus).

During enclave initialization, the CPU \emph{measures} the enclave's code and configuration, which constitute the enclave's identity (i.e.\ its \texttt{MRENCLAVE} value).
The enclave can \emph{seal} data by encrypting it with a CPU-protected key that can only be accessed from enclaves running on the same CPU with precisely the same \texttt{MRENCLAVE} value.

\subsubsection{SGX Remote Attestation}
\label{sec:preliminaries-sgx-attestation}

Remote attestation is the process through which one party, the \emph{verifier}, can ascertain the precise hardware and software configuration of a remote party, the \emph{prover}.  
The objective is to provide the verifier with sufficient information to make a trust decision about the prover.

SGX supports remote attestation by having verifiers rely on the \emph{Intel Attestation Service} (IAS) to validate attestation evidence produced by a remote prover enclave~\cite{Anati2013}. 
The SGX SDK includes a pre-defined set of functions to run a remote attestation protocol.
This default protocol is designed for a service provider to provision secrets to an enclave, and as such, the service provider is authenticated by the enclave as part of the protocol.

The protocol starts with the verifier sending a challenge to the enclave.
The enclave generates an ephemeral Diffie-Hellman (DH) key pair and sends the public key to the verifier.
The verifier contacts the IAS to retrieve a \emph{signature revocation list (SigRL)}, which is needed to prove that the platform has not been revoked.
The SigRL does not have to be fetched in every run of the protocol, but must be sufficiently recent for the attestation to be verified.
The verifier generates its own DH key pair and sends the public key to the enclave, signed with the verifier's long-term key.
The verifier's long-term public key is hardcoded into the enclave to allow the enclave to authenticate the verifier. 
Thus the enclave should be aware of all verifiers that will be able to attest the enclave.
SGX then produces a \emph{quote} of the enclave, which includes a hash of both public keys and a key derived from the common secret.
In this way the quote unambiguously identifies the enclave and cryptographically binds the public key to this enclave.
Finally, the quote is sent to the verifier, who uses the IAS to validate the quote.
The response from the IAS is signed by an Intel public key.

\section{System Model and Requirements}
\label{sec:system_model}

\subsection{Overview}

\begin{figure}[ht]
	\begin{center}
    \includegraphics[width=1\columnwidth]{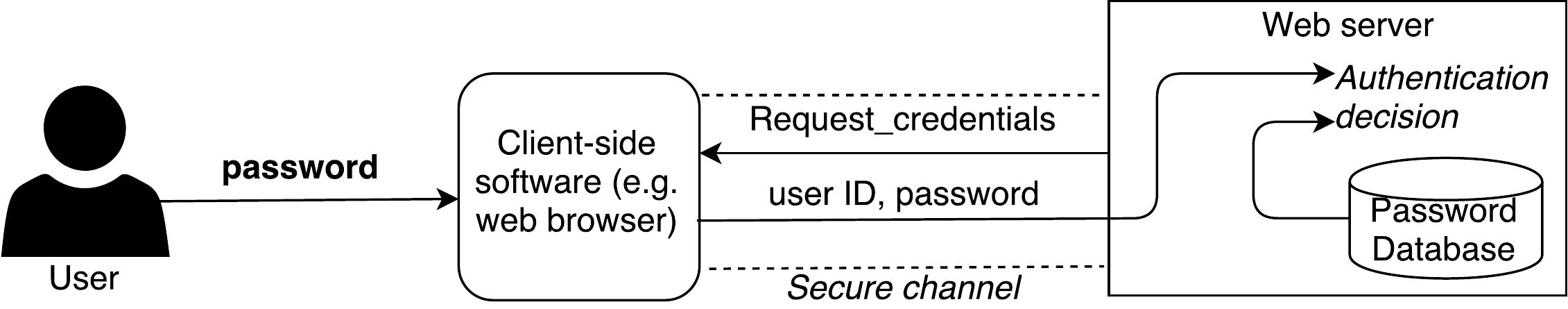}
    \caption{Generalized model of a password-based authentication system.}
    \label{fig:system_model}
	\end{center}
\end{figure}

We use the term \emph{password} to refer to any user-memorized authentication secret.
As explained in Section~\ref{sec:preliminaries}, passwords are generally weak secrets (e.g.\ 20~bits of entropy~\cite{Bonneau2012}) that are re-used across multiple services.
Figure~\ref{fig:system_model} is a generalized model of a password-based authentication system.
With the assistance of client-side software (e.g.\ a web browser), a user sends her user ID and password to a server over a secure channel (TLS), and based on this, the server makes an authentication decision.
Although real systems are undoubtedly more complicated, they can all be logically represented as the model in Figure~\ref{fig:system_model}.  
Therefore, for clarity of exposition and without loss of generality, we use this model and terminology throughout this paper. 

Confidentiality of user passwords may be compromised in multiple ways, including: 
\begin{itemize}
\item Users disclose a password directly to the adversary under the mistaken impression that the password is being sent to the intended service (e.g.\ phishing attacks).
\item Information about the password, or the password itself, is leaked from the server (e.g.\ stolen password database).
\item Compromise of the server itself (e.g.\ a web server memory snooping).
\end{itemize}

Any comprehensive solution for protecting passwords must defend against \emph{all} these different attack avenues.

\subsection{Adversary Model}
\label{sec:system_model-adversary}

The goal of our adversary is to learn users' plaintext passwords.
Our adversary model is stronger than that of previous solutions -- we allow for the possibility that the adversary may have the capabilities of the \emph{operator} (owner/administrator) of the web server itself.
This covers both compromised web servers, as well as malicious server operators (for clarity of exposition, we use the term \emph{rogue server} to refer to both of these).
The widespread practice of reusing the same or similar passwords across services provides the incentive for the adversary to learn plaintext passwords, which they can then abuse in different ways, including (i) masquerading as the user on a \emph{different} service, where the user may have used the same or similar password; (ii) accessing user data encrypted by a password-derived key (e.g.\ for encrypted cloud storage services, especially, if forced legally or illegally); and (iii) selling/leaking user passwords to other malicious entities.

Concretely, we allow the adversary the following capabilities, covering rogue servers and weaker external adversaries:

\textbf{Access password database:} the adversary has unrestricted access to the database of stored passwords.
This models both rogue servers and weaker external adversaries who may steal the password database.

\textbf{Modify web content:} the adversary can arbitrarily modify the content sent to the user (including active content such as client-side scripts).
This also models weaker external adversaries who can modify content using attacks such as cross-site scripting (XSS).

\textbf{Access to server-client communication:} the adversary can read all content sent to the web server, including content encrypted by a TLS session key. 
For a rogue server, such access is easily obtained, and for external network adversaries, such access could be obtained through successful attacks against TLS (e.g.~\cite{rfc7457,Aviram2016}).
 
\textbf{Execute server-side code:} the adversary has full knowledge of all software running on the server, and is able to execute arbitrary software on the server.
This captures a powerful attacker who successfully gains access and escalates privileges on the server, or a malicious server operator.

\textbf{Launch phishing attacks:} the adversary can launch state-of-the-art phishing attacks, including targeted attacks.

We assume that the adversary \emph{cannot} compromise the client-side software, including the user's OS and browser, but the adversary can send any content to the client.
Although client-side security is important, it is an orthogonal problem and is thus out of the scope of this work.
It is addressed by other means, such as client-side platform security, use of anti-malware tools etc.
Our focus in this paper is how password-based authentication systems can be made resilient against strong server-side adversaries.

We assume that the adversary is computationally bounded, and thus cannot feasibly subvert correctly-implemented cryptographic primitives.
We also assume that the adversary does not have sufficient resources to subvert the security guarantees of hardware-based TEEs through direct physical attacks.
We discuss side-channel attacks specific to SGX in Section~\ref{sec:evaluation-security}, including some counter-measures.
Denial of service (DoS) attacks are out of scope because a rogue server can always mount a DoS attack by simply refusing to respond to requests.

\subsection{Requirements and Objectives}
\label{sec:system_model-requirements}

Given our strong adversary model, a full server compromise could undoubtedly cause significant damage (e.g.\ theft, loss, or modification of user information).
Our aim is to guarantee that even such a compromise will not leak users' passwords.
This is a valuable security guarantee because passwords are re-used across multiple services, and sometimes used to derive keys to encrypt content stored in the cloud.  

Guaranteeing password confidentiality mitigates attacks on the server and password databases, but not phishing attacks. 
Thus we also need to provide users with the means to easily and effectively determine when it is safe to enter their passwords.
Therefore, we define the following two formal requirements for a comprehensive solution for protecting a password-based authentication system:  

\begin{enumerate}[label={R\theenumi},leftmargin=*,labelindent=0mm,labelsep=2mm]

\item \label{R1} \textbf{Password protection:} The server must protect users' passwords by fulfilling all the following criteria:
\begin{enumerate}[label=(\roman*),leftmargin=4mm,labelindent=0mm,labelsep=1mm]
\item The strong adversary defined above cannot obtain users' passwords through any means other than guessing (e.g.\ he cannot observe passwords in transit or while they are being processed on the server). 
\item \emph{Offline} password guessing must be computationally infeasible, irrespective of the strength of the password.
\item \emph{Online} password guessing must be throttled, irrespective of the adversary's computational capabilities.
\end{enumerate}

\item \label{R2} \textbf{User awareness:} End users must be able to easily and accurately determine whether it is safe to enter their passwords when prompted by a given server (i.e.\ indirectly determine whether the server fulfils Requirement~\ref{R1}).

\end{enumerate}

Note that Requirement~\ref{R2} \emph{does not} mandate users to understand the precise technical security guarantees of Requirement~\ref{R1}, but rather that the solution should enable users to determine \emph{which} servers meet this requirement, and will thus protect passwords.

In order to be effective, any solution for protecting passwords must be deployable in real-world systems.
Therefore, in addition to the above two security requirements, we also target the following deployment objectives:

\textbf{Minimal performance overhead:} The solution should not noticeably degrade the performance of password-based authentication systems, either in terms of \emph{latency} (the time required to complete a single authentication attempt), or \emph{scalability} (the overall rate at which authentication attempts can be evaluated).

\textbf{Minimal software changes:} It should be possible to integrate the solution into a wide range of existing software systems without requiring significant effort.

\textbf{Ease of upgrade:} It should be possible to upgrade existing password-based authentication systems to adopt the solution transparently (e.g.\ without immediately mandating users to reset their passwords). 
Existing mechanisms for changing/resetting passwords should also remain unaffected.

\textbf{Backup and recovery:} The solution should include a secure backup and recovery mechanism to ensure that it can continue operating in the event of (multiple) hardware failures.

\section{Design}
\label{sec:design}

\begin{figure}[ht]
	\begin{center}
    \includegraphics[width=1\columnwidth]{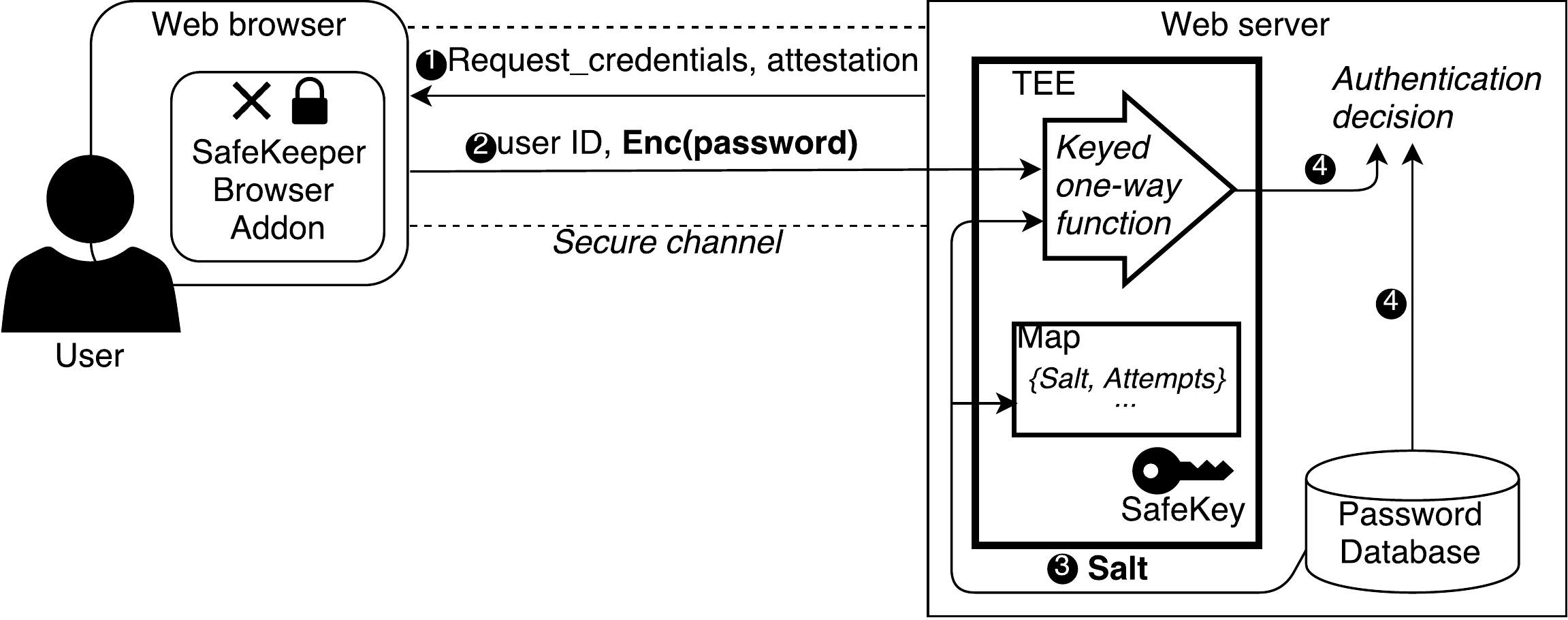}
    \caption{Overview of the \SafeKeeperCaption design.}
    \label{fig:overview}
	\end{center}
\end{figure}

As shown in Figure~\ref{fig:overview}, \SafeKeeper consists of a \emph{server-side password protection service}, which computes a \emph{keyed} one-way function on passwords before they are stored in the database.
An adversary must obtain the key (i.e.\ pepper) used in this one-way function to launch feasible offline guessing attacks against a stolen password database.
This key is randomly generated, and of sufficient length (e.g.\ 128 bits) to defeat brute-force attacks.
We refer to this key as the \emph{\SafeKey}, since it is generated and protected by a server-side Trusted Application (TA, executing within the TEE); see Section~\ref{sec:design-tee}.

With direct access to the password protection service, an untrusted server operator or an adversary who has compromised the server can also perform online password guessing attacks.
In this case, the adversary supplies a guessed password, and the password protection service returns the processed result, which the adversary can compare with the stored password value.
To defend against this attack, the password protection service must limit the rate at which it processes passwords.
\SafeKeeper achieves this by enforcing rate limiting in the TEE; see Section~\ref{sec:design-rate_limiting}.

Furthermore, the rogue server may attempt to observe passwords \emph{before} they are sent to the password protection service.
A secure channel between the web browser addon and the server (e.g.\ a TLS connection) is insufficient as the server-end of such a connection is controlled by the server operator. Instead, \SafeKeeper establishes a separate secure channel directly between the browser and the TEE-protected password protection service; see Section~\ref{sec:design-attestation}.
Finally, the browser needs some way to determine which input data should be sent via this secure channel to the password protection service (e.g.\ passwords but not user IDs).
To improve usability, the server operator defines which input fields will be protected, and then the \SafeKeeper browser addon displays this information to the user.
The user is thus only required to validate that the password field is protected; see Section~\ref{sec:design-ui}.

\subsection{Server-side Password Protection}
\label{sec:design-tee}
The \SafeKeeper password protection service (\SafeKeeper TA) is designed to be a drop-in replacement for existing one-way functions (e.g.\ password hash algorithms).
As such, it takes as input the concatenation of the password and the corresponding salt value, and outputs the result of a keyed one-way function, which the server stores in its database.
To protect the \SafeKey used in this function, \SafeKeeper computes the one-way function inside a server-side TEE.
The TEE provides strong isolation (e.g.\ hardware-enforced) from all other software on the server (including the OS/hypervisor), and thus ensures that the \SafeKey is available only to the TA code.
Even if an adversary obtains the password database, he cannot perform an \emph{offline} password guessing attack because he would need the \SafeKey to test his guesses against the database.
Since offline attacks are no longer possible, \SafeKeeper can use any cryptographically secure one-way function, i.e.\ specifically-designed password hash functions (e.g.\ Argon2~\cite{Argon2}) are not essential (but can be used).
The adversary is thus forced to try \emph{online} guessing attacks, which are
mitigated by \SafeKeeper's rate-limiting mechanism.

\subsection{Rate Limiting}
\label{sec:design-rate_limiting}
Ideally, the password protection service should perform rate limiting on a per-account basis to: (i) protect each account password, and (ii) avoid rate-limiting a user due to the actions of other users. 
Note that, in order to serve as a drop-in replacement in existing authentication frameworks, the \SafeKeeper TA only computes the keyed one-way function, but does not make or learn the authentication decision.
In other words, it cannot distinguish between the two scenarios: (i) when the user account is under guessing attack, and (ii) when a legitimate user is attempting to login multiple times within a short period of time (with or without a valid password). 
Also, as the function does not take the user ID as input, we cannot implement user ID based rate limiting.
Changing the function to include user IDs as input would require non-trivial changes to the server software, limiting \SafeKeeper's use as a drop-in replacement.
As a work-around, we rate limit each account using the unique per-account salt values, which are provided to the one-way function as part of regular operation.
It is general security best-practice to use unique salt values for each account, but if a server operator chooses the same salt for multiple (or all) accounts, he only restricts his own guessing capability.
Furthermore, since the salt is a fixed pre-determined length, the adversary cannot perform a type-substitution attack by providing a salt of a different length, e.g.\ concatenating the first few characters of a guessed password with the salt value.

\SafeKeeper limits password processing for each account (salt) based on a \emph{quantized maximum rate}.
Simply enforcing a \emph{maximum rate} (e.g.\ waiting 10 minutes between password attempts) would negatively impact usability in cases where the user mistyped a password.  Instead, the quantized maximum rate allows a fixed number of attempts within a pre-defined time interval, but doesn't mandate a delay between these attempts.
For example, each user could be allowed 144 attempts that could be used at any time within each 24-hour period.
After exhausting these attempts, the user has to wait until the following time period, when the count is reset.
Overall, this results in the same rate as waiting 10 minutes between guesses, but significantly improves usability when multiple attempts are required in quick succession.

\subsection{Remote Attestation}
\label{sec:design-attestation}
To securely transmit a user password, the \SafeKeeper browser addon must correctly authenticate the \SafeKeeper password protection service (\SafeKeeper TA) running inside a TEE, via remote attestation.
In addition, the addon must establish a secure channel directly with the TA.
\SafeKeeper uses binary attestation to assure the addon of the precise TA running inside the server-side TEE.
To verify that it is communicating with a genuine \SafeKeeper TA, the addon verifies the quote, and then checks the binary measurement against a whitelist of known \SafeKeeper TAs.
Since the same TA can be used by many websites, and the functionality of the TA is unlikely to change, the whitelist of genuine \SafeKeeper TAs will be short, and can be built into the browser.
The attestation protocol includes a key agreement step through which the browser and the TA establish a shared session key.
Note that the attestation protocol provides only unilateral authentication of the \SafeKeeper TA towards the client; i.e.\ the client software or the user is not authenticated during attestation.
Thus, anyone, including an adversary can establish a connection and interact with the TA.
However, since the key agreement step is cryptographically bound to the TA's remote attestation, the adversary cannot perform a man-in-the-middle attack when legitimate clients communicate with the TA.

Using the shared session key, the browser encrypts the password before the page is submitted.
The encrypted password is sent to the server in place of the original password, and may be wrapped in additional layers of encryption (e.g.\ TLS).
On the server, the encrypted password is input to the TA, which decrypts it using the shared key and performs the cryptographic one-way function.

\subsection{Client-side Assurance Mechanism}
\label{sec:design-ui}
\SafeKeeper's client-side assurance mechanism can be added to existing web browsers e.g.\ by installing a browser addon.
This addon executes the remote attestation protocol and establishes a secure channel with the TA.
If the attestation succeeds, the addon changes its appearance to signal this to the user (e.g.\ an icon in the browser toolbar).

\begin{figure}[t]
	\begin{center}
    \includegraphics[width=1\columnwidth]{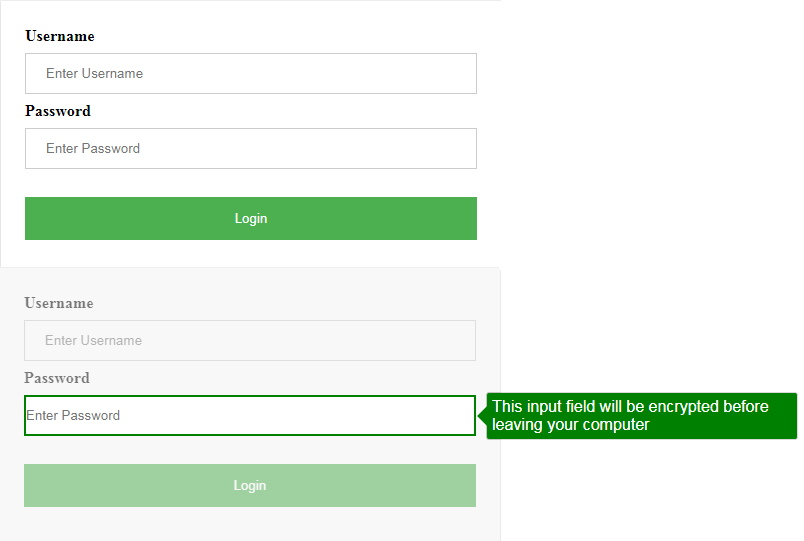}
    \caption{\emph{Top:} The original webpage; \emph{Bottom:} \SafeKeeperCaption highlighting the password input field.}
    \label{fig:highlighting}
	\end{center}
\end{figure}

The server specifies in the web page which input fields should be encrypted and sent to the \SafeKeeper password protection TA.
The addon parses this information and encrypts any text entered into these fields. 
However, a rogue server may specify that some non-password fields should be protected, while the actual password field is not protected.
To prevent this, the \SafeKeeper addon signals to the user which input fields are protected by greying out the whole page, highlighting only the text input fields it will encrypt, and displaying an information tooltip, as shown in Figure~\ref{fig:highlighting}.

An alternative approach would be to ask the user to designate certain text as passwords (e.g.\ by requiring the user to type ``\texttt{@@}'' as in PwdHash~\cite{Ross2005}).  However, our approach was chosen in order to reduce the cognitive load on users, since they are required only to \emph{check} that the relevant input field is protected, and may also prevent errors where users designate an incorrect field to be protected (e.g.\ user ID).

Additionally, the adversary could attempt to spoof the highlighting performed by \SafeKeeper (e.g.\ by highlighting fields that are not actually protected).
To mitigate this, we use a similar principle to a \emph{secure attention sequence} by requiring the user to click on the browser addon icon to activate the highlighting.
This click cannot be detected or prevented by the adversary (as it is outside of the browser DOM).
After the user has clicked, the \SafeKeeper icon is again changed to indicate that it is in the \emph{highlighting mode}.
This provides a spoofing-resistant mechanism for signaling to the user which input fields will be protected.
The user is thus assured that any password entered into this type of protected input field will always be protected by \SafeKeeper, regardless of the identity of the website or the behavior of the server. 

Unlike many other client-side approaches (e.g.\ password managers), the \SafeKeeper browser addon is stateless and user-agnostic.
This makes our client-side assurance mechanism a good candidate to be integrated into future versions of web browsers.
Full details of our implementation of this mechanism are discussed in Section~\ref{sec:implementation-ui} and evaluated in Section~\ref{sec:evaluation-usability}.

\section{Implementation}
\label{sec:implementation}

We have implemented a fully-functional open-source prototype of \SafeKeeper.
In this section, we describe the specific implementation challenges and our solutions.

\subsection{Server-side Password Protection}
\label{sec:implementation-tee}

Our implementation of the server-side password protection service uses Intel's recent Software Guard Extensions (SGX).
However, \SafeKeeper can use any equivalent TEE that provides isolated execution, sealed storage, and remote attestation.
We use SGX for its ease of use and superior performance, and increasing prevalence on server platforms (e.g.\ Intel Xeon).

\begin{lstlisting}[caption={Password protection service enclave
interface},captionpos=b, float, frame=tblr, label=lst:enclave]
  init(in_sealed(key||attempts));
  process(in(password||salt), out(CMAC));
  reset_attempts();
  shutdown(out_sealed(key||attempts));
\end{lstlisting}

The design of our SGX enclave is kept minimalistic, consisting of only four \texttt{ecalls}; see Listing \ref{lst:enclave}.
When the enclave is started for the first time, the \texttt{init()} uses Intel's true random number generator (via the \texttt{RDRAND} instruction), to generate a new strong random \SafeKey.
When the enclave is later restarted, this function is used to pass previously-sealed data to the enclave.
The \texttt{process()} function calculates the keyed one-way function on the password using the \SafeKey and returns the result.
We use the Rijndael-128 CMAC function, as this meets our security requirements and can be computed using the AES-NI hardware extensions.
The \texttt{reset\_attempts()} function forms part of our in-enclave rate-limiting mechanism.
The \texttt{shutdown()} function is used to perform a graceful shutdown of the enclave (e.g.\ in case the server needs to reboot).
This function seals the \SafeKey and the current state of the enclave.

We integrate \SafeKeeper's password protection service into the \texttt{PHPass} library, which is widely used for password hashing in popular web platforms including WordPress, Joomla, and Drupal.  
By default, \texttt{PHPass} uses a software implementation of MD5 with 256 iterations.
We replaced this with a single call to our enclave, using the PHP-CPP framework\footnote{\url{http://www.php-cpp.com}} to invoke C/C++ enclave functions from PHP.

\subsubsection{Rate limiting}
\label{sec:implementation-rate_limiting}

Apart from the web server's rate limiting (e.g.\ the use of Captchas after a certain number of failed attempts), we implement the rate-limiting algorithm (Section~\ref{sec:design-rate_limiting}) within the TEE-protected password service. 
Our enclave program keeps an in-memory map (using C++ \texttt{std::map}) that associates each salt ($i$) with a number of remaining attempts ($attempts_i$).
To provide maximum flexibility, our implementation uses a 64-bit salt and a 32-bit integer for $attempts_i$, although this can be decreased to reduce memory consumption if needed.

When \texttt{process()} is called for salt $i$, this function first checks the value of $attempts_i$; if the value is zero, the function returns only an error; otherwise $attempts_i$ is decremented by one and the CMAC result is returned.
The enclave stores $t_{reset}$, the time at which all $attempts$ values are reset to a predefined value, $attempts_{max}$.
When \texttt{reset\_attempts()} is called, this function first obtains the current time; then, if $t_{reset}$ has passed, sets all $attempts$ values to $attempts_{max}$ and increases $t_{reset}$ by a predefined value.

Although the effective rate can be set by the enclave developer, it is verified by the user via the browser addon, so a malicious developer cannot set an arbitrarily high rate.
In our implementation, we use 144 attempts per 24 hours.
Operations involving time make use of the SGX trusted time API: \texttt{sgx\_get\_trusted\_time()}; when called, the platform returns the current time (with second precision) and a platform-generated nonce.
Any change in the nonce value between two invocations of this function indicates that there may have been an attack on the platform, and thus the returned time value should not be trusted.
During initiation, the \SafeKeeper TA obtains and stores this nonce in its protected memory.

To allow the enclave to be restarted (e.g.\ if the server is rebooted), the \texttt{shutdown()} function is used to securely store the state information outside the enclave.  Specifically, the enclave seals the \SafeKey, the map of salts and $attempts$ values, $t_{reset}$, and the time nonce.
This sealed data can be restored to the enclave via the \texttt{init()} function.
The enclave uses monotonic counters\footnote{\url{https://software.intel.com/en-us/node/709160}} to prevent rollback attacks in which the adversary attempts to restore old
sealed data.
When the enclave starts, the \texttt{init()} function increments a hardware-backed monotonic counter.
The \texttt{shutdown()} function seals this counter value along with the enclave's state.
When the enclave is restarted, this counter value is checked to ensure that the latest sealed state is being restored.

A rogue server may attempt to reset the $attempts$ values by abruptly killing the enclave without first sealing its state.
However, the enclave will detect this because the counter value in the sealed data will not match the current value of the hardware monotonic counter.
In this case, the enclave has no way of restoring the previous $attempts$ values -- the data has been irreversibly lost.
Therefore, the only secure course of action is to set $t_{reset}$ to some predetermined time in the future, and set all $attempts$ to zero (i.e.\ to impose the maximum penalty).
This captures the worst-case scenario in which the adversary had exhausted all guessing attempts against all accounts.
Note that, during normal operations, enclave crashes should be rare, and with proper load balancing the effects of abnormal enclave crashes can be amortized.

\subsubsection{0-round-trip attestation}
\label{sec:implementation-attestation}

Remote attestation is used to assure the browser that it is communicating with a genuine \SafeKeeper password protection service running inside an SGX enclave.
It is achieved by obtaining a \emph{quote} from the enclave and verifying it using the IAS.
The quote includes an unforgeable representation of the code executing inside the enclave.
As described in Section~\ref{sec:preliminaries}, the remote attestation protocol provided with the SGX SDK is designed to provision secrets to an enclave from a trusted source, and thus requires mutual authentication.
This protocol involves four messages and two round-trips.
In \SafeKeeper there is no requirement for the browser addon to authenticate itself to the enclave.
Any entity can request a quote from the enclave and then decide whether to establish a secure channel.
We thus modify the attestation protocol to optimize it for \SafeKeeper.

We denote our optimized protocol as \emph{0-round-trip} attestation because it does not require any round-trip message exchanges between the client and server.
In our protocol, the SGX signature revocation list (SigRL) is fetched by the \SafeKeeper password service (i.e.\ by the server, instead of the client).
Our implementation fetches an updated list once per hour.
Our enclave generates a Diffie-Hellman (DH) key-pair and obtains an SGX quote containing the hash of the DH public key. 
This binds the public key to the enclave's identity (i.e.\ its \texttt{MRENCLAVE} value).
When a web page is requested by a browser, the server includes the quote as an HTTP-header in the response.
The client then verifies the quote using the IAS (see Section~\ref{sec:implementation-detecting}).

Since the enclave prepares and obtains the quote \emph{before} any client connection, there is no overhead of generating a quote on-demand for each client.
Unlike the original protocol, our protocol does not require the client to supply a challenge to be included in the quote.
This is because the current version of SGX measures the enclave's code when the enclave is initialized and then does not allow the code to be changed.
Therefore, even without a challenge, the verifier is assured that the public key included in the quote is fresh and belongs to the enclave.

Also unlike the original protocol, in our modified version, the enclave does not generate a new public key for each client.
Since each client generates its own DH key-pair, the resulting keys used to encrypt the passwords will still be completely different for each client (i.e.\ no client will be able to read another's password).
One disadvantage of this approach is that a compromise of the enclave's single public key could potentially reveal all encrypted passwords.
However, this could only be achieved by subverting the security guarantees of SGX, in which case the \SafeKey could also be compromised and used in offline password guessing attacks.
As explained in Section~\ref{sec:system_model}, adversaries capable of subverting hardware-enforced TEEs are out of scope of this work.

\subsubsection{SGX remote attestation proxy}
\label{sec:implementation-proxy}

The browser addon cannot check the quote directly because all communication with the IAS must be performed via a mutually-authenticated TLS channel, using a
TLS client certificate that the developer has registered with Intel.\footnote{\url{https://software.intel.com/en-us/articles/certificate-requirements-for-intel-attestation-services}}
We assume everyday users will be unable or unwilling to be registered with Intel for \SafeKeeper or any such services.
To overcome this challenge, we  implement an \emph{SGX remote attestation proxy}.
Our proxy server accepts unilaterally-authenticated TLS connections from verifiers (e.g.\ the \SafeKeeper browser addon) and forwards the contents to the IAS via the required
mutually-authenticated TLS channel.
The results from the IAS are returned to the callers.
Since the IAS signs its response message and provides a certificate chain rooted in a public Intel verification certificate, the verifiers \emph{do not} need to trust our IAS proxy.
We implemented this proxy using an Nginx\footnote{\url{https://nginx.org/en/}} server configured as a reverse HTTP proxy with TLS support on both the upstream server and the incoming connections.
The TLS certificate of the proxy is pinned in the \SafeKeeper browser addon.
Since the enclave uses the same public key and quote for all clients, our proxy can cache the quote verification response from IAS to reduce the verification latency.

\subsection{Client-side Assurance Mechanism}
\label{sec:implementation-ui}

We implement \SafeKeeper's client-side assurance mechanism as an addon for the Google Chrome browser (similar implementations can also be developed for the other browsers).
We assume that users can download and install this addon securely (e.g.\ using the Chrome Web Store), and receive software updates when available.
Note that browser vendors are actively working to ensure the security of browser addons.\footnote{\url{https://blog.chromium.org/2015/05/continuing-to-protect-chrome-users-from.html}}

Our browser addon is written in JavaScript and consists of two parts: (i) a \emph{background script} implementing the main functionality; and (ii) a \emph{content script}  injected into each web page in order to interact with the page content.
The addon therefore requires permission to access tabs data, use the browser storage, and capture and modify certain web requests.

\subsubsection{Detecting the password protection service}
\label{sec:implementation-detecting}

\begin{figure}[h]
	\begin{center}
    \includegraphics[width=0.8\columnwidth]{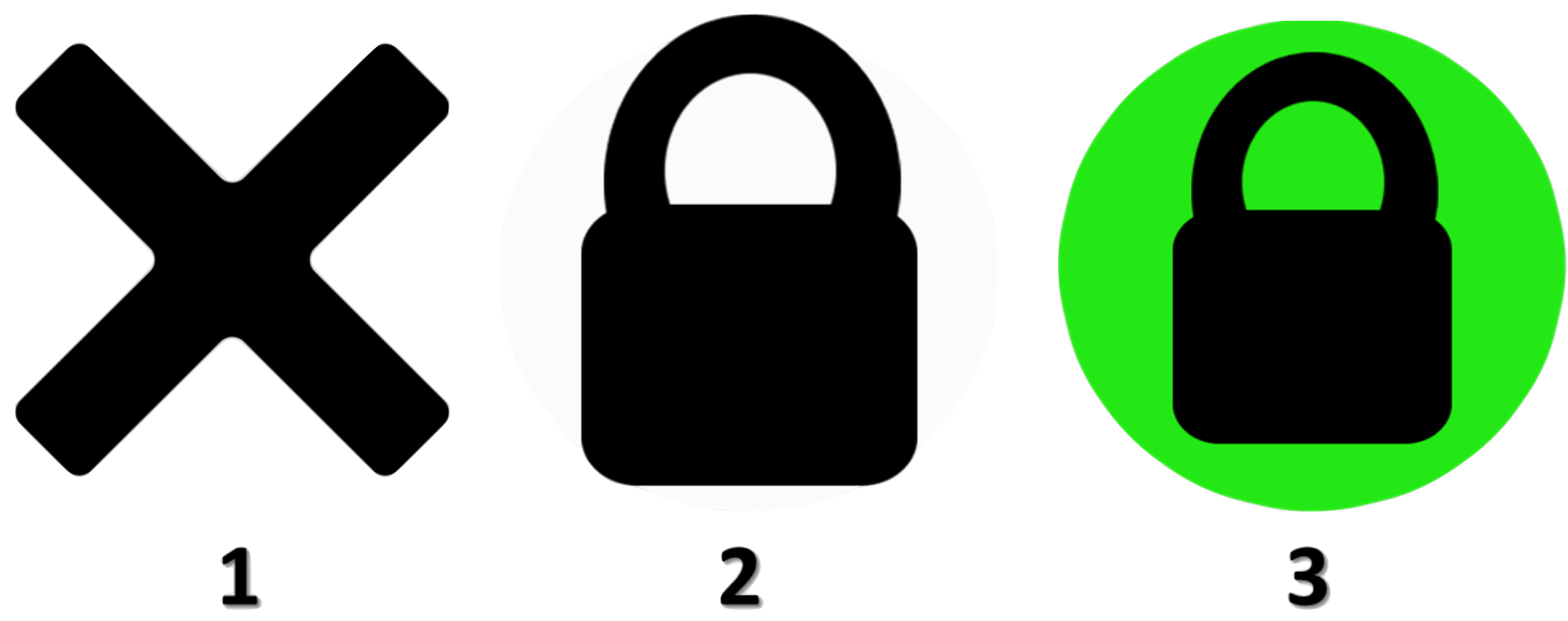}
    \caption{\SafeKeeperCaption browser addon icons: the first one indicates that \SafeKeeperCaption is unavailable on the website, or that the attestation protocol has failed; the second icon shows that \SafeKeeperCaption is supported and that a secure channel has been established; and the third icon is used when \SafeKeeperCaption is highlighting the protected input fields (see Section~\ref{sec:implementation-highlighting}).}
    \label{fig:icons}
	\end{center}
\end{figure}

Web servers with \SafeKeeper support will send an SGX quote in the HTTP response header of web pages with protected fields (e.g.\ the login form).
This quote is processed by the addon's background script, which verifies the integrity of the service and the validity of the quote.
Specifically, the addon extracts from the quote the unique hash representing the enclave's loaded code (i.e.\ the \texttt{MRENCLAVE} value), and checks if this is included in its list of trusted values.
This list can be updated in a similar manner to updating the browser.
If the enclave's identity is trusted, the addon sends this quote to the IAS, via our IAS proxy (see Section~\ref{sec:implementation-proxy}).
Upon receiving a response, the addon checks the result from the IAS and verifies the signature on the response using Intel's public IAS verification key.
If the attestation process is successful, the background script generates a new DH key pair for the website, and establishes the shared key using the enclave's public
key.
This is stored in a local variable and also sent to the content script.
Finally, the background script changes the icon of the addon to indicate the website's support of \SafeKeeper; see Figure~\ref{fig:icons}.

\subsubsection{Highlighting protected input fields}
\label{sec:implementation-highlighting}

When a page is loaded, the injected content script checks for the \texttt{\SafeKeeper} metatag.
If present, this tag specifies which input fields must be encrypted by \SafeKeeper.
The content script stores this information in a local variable.

\begin{figure}[ht]
	\begin{center}
    \includegraphics[width=0.7\columnwidth]{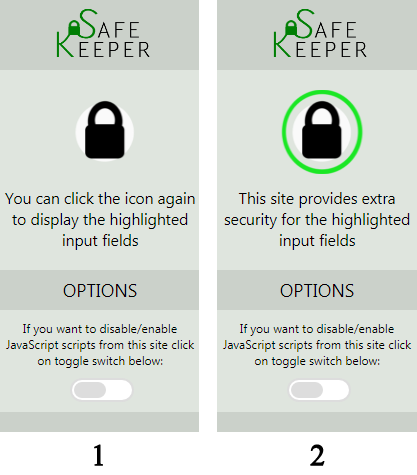}
    \caption{\SafeKeeperCaption popup windows: (1) the page supports \SafeKeeperCaption, but the protected input fields are not currently highlighted; (2) \SafeKeeperCaption is highlighting the protected fields in the page.}
    \label{fig:popups}
	\end{center}
\end{figure}

When the user clicks the browser addon icon, a popup window appears to provide information about the current web page; see Figure~\ref{fig:popups}.
Clicking the addon icon also serves as the secure attention sequence described in Section~\ref{sec:design-ui}, as this cannot be manipulated by the adversary-controlled web page script.
This action sends a message to the content script, which modifies the website's DOM elements to highlight the protected input fields.
Specifically, the content script clones the protected input fields, highlights them in green, and attaches an information tooltip to inform the user why a field is highlighted.
It then decreases the opacity of all the other elements in the website's body section to ensure that the highlighted fields stand out to the user.
Cloning the protected input fields is necessary as it is not allowed to specify a different opacity for nested elements within the body section.
When the user clicks on the icon again, the content script increases the opacity value for all the body elements, copies any input from the cloned fields back to the original, and removes the clones.
When the web page is submitted, the content script encrypts all values from the protected input fields using the shared key agreed with the \SafeKeeper password service.

\subsubsection{Defending against malicious client-side scripts}
\label{sec:implementation-scripts}

As defined in our adversary model (Section~\ref{sec:system_model}), a malicious server operator has the capability to modify the content of the web page, including adding client-side scripts.  This poses several threats.  
First, as client-side scripts can read data from any field on the page, a malicious script may attempt to read the password as it is typed by the user.  
Client-side scripts that exhibit this behaviour for other types of personal information have already been observed in the wild.\footnote{\url{https://gizmodo.com/before-you-hit-submit-this-company-has-already-logge-1795906081}}
With the exception of disabling the malicious script, there is no way to avoid this using current browser addon technologies since the script is executing in the same domain
as the text input field.
Second, malicious client-side scripts can perform the same type of highlighting of input fields as the \SafeKeeper browser addon.
For example, the adversary could use this to highlight input fields that are not actually protected.
Third, although the adversary cannot detect when the \SafeKeeper highlighting has been activated, he can add a time delay to his malicious script (e.g.\ 10 seconds).
His aim is to make his malicious script run \emph{after} the user has clicked the \SafeKeeper icon to activate highlighting mode, in which case the malicious script will change the highlighted fields while \SafeKeeper is in highlighting mode.

To mitigate these attacks, \SafeKeeper provides a mechanism for disabling all other client-side scripts on a particular website.  Currently Google Chrome does
not provide a direct option to disable client-side scripts for particular websites, so we have developed an alternative approach to provide this
functionality using a \emph{Content Security Policy} (CSP).
Using the toggle switch on the \SafeKeeper popup window, users can disable scripts for individual websites. 
\SafeKeeper then reloads the page and injects our custom CSP metatag into the header.
This CSP still allows our injected content script to run, but prevents all other scripts from executing on the page.
A limitation of this approach is that it leads to a race condition between the time our CSP is injected, and the loading of any other scripts.
However, our experimental evaluation showed that only pages loaded directly from \texttt{localhost} were fast enough to evade our CSP.
When scripts are disabled for a particular website, the addon stores this information in Chrome's local storage and continues to disable scripts on future visits to this website, until the user re-enables scripts.

However, disabling client-side scripts often negatively affects the usability of the web page.
Therefore, by default \SafeKeeper allows client-side scripts.
Careful users can still turn-off all client-side scripts manually when needed, e.g.\ if they are unsure about the server's bona fides. 
In Section~\ref{sec:extensions}, we present an alternative approach that ensures the same level of security without disabling client-side scripts.

\section{Evaluation}
\label{sec:evaluation}

We evaluated the security, usability, performance, and deployability of \SafeKeeper.

\subsection{Security Analysis}
\label{sec:evaluation-security}

As defined in Requirement~\ref{R1} (Section~\ref{sec:system_model-requirements}), a comprehensive solution for protecting passwords must (i) prevent even the strongest adversary (i.e.\ a rogue server) from observing passwords in transit or while they are being processed on the server; (ii) prevent offline password guessing attacks; and (iii) throttle online password guessing, irrespective of the adversary's computational capabilities.
We analyse the security guarantees provided by \SafeKeeper against each of these classes of attacks, and against other attacks that may arise from the technologies used in our implementation.

\subsubsection{Passwords in transit and on the server}
The secure channel, based on the DH key agreement between the browser addon and the enclave, ensures the confidentiality of passwords while in transit over the network, and while they are being processed on the server, before they are input to the enclave.
Since this channel is cryptographically bound to the enclave's remote attestation quote, the client-side software is assured that it is communicating with the correct enclave. Note that, this channel's security is independent of any other layer, such as TLS, established between the browser and web server.
As such, password confidentiality remains unaffected even if there are flaws in TLS/HTTPS protocols (e.g.~\cite{rfc7457,Aviram2016}).

\subsubsection{Offline guessing}
In an offline attack, the adversary would attempt to guess the password and test his guess directly against the password database (e.g.\ leaked password hash databases), bypassing any online guessing prevention by the web server.
The adversary is therefore not subject to rate-limiting, and can perform guesses at the maximum rate supported by his available hardware.
However, to test these guesses, the adversary must also guess the \SafeKey used in the CMAC calculation, and thus making offline attacks infeasible even against very weak passwords.

\subsubsection{Online guessing}
To avoid having to guess the \SafeKey, a malicious server operator could perform an \emph{online} attack by submitting password guesses to the \SafeKeeper password protection service, running in the SGX enclave.
However, the adversary's rate of guessing is then constrained by \SafeKeeper's rate-limiting mechanism described in Section~\ref{sec:implementation-rate_limiting}.
This effectively increases the difficulty of guessing the password by increasing the time required. 
For example, an average strength password is estimated to provide approximately 20~bits of entropy~\cite{Bonneau2012}, so if \SafeKeeper's effective maximum rate is set to 144 authentication attempts per day, the average time required to guess this password (i.e.\ $2^{19}$ guesses) would be approximately 10 years.
Note that even this type of rate limiting cannot protect very weak passwords for long (e.g.\ short passwords or those that appear in lists of frequently used passwords).
This is a fundamental limitation of password-based authentication.

Even though our rate limiting is based on the salt, which is under the adversary's control, the adversary cannot increase his guessing rate by manipulating the choice of salts.
For example, if the adversary chooses the same salt for all accounts, he can test his guesses against all passwords in the database (e.g.\ 144 guesses per day, tested against all accounts).
However, if he chooses unique salts he can test the same number of guesses against each individual account, and possibly vary guesses between accounts (e.g.\ 144 guesses per day against each account). Additionally, he may try to add fake accounts with guessed passwords to leverage the fact that the same passwords will result in the same CMAC values, if a global salt is used. However, as passwords of new accounts must also be processed by the \SafeKeeper password protection service, the number of guessing attempts per day remains unaffected. 
An external adversary (i.e.\ without compromising the web server) will also be restricted by any online guessing prevention mechanisms (e.g.\ Captcha) implemented by an honest server operator.

\subsubsection{Implementation-specific attacks}

In addition to the main classes of attacks discussed above, we also analyse attacks that may arise as a result of the specific technologies used in our implementation.

\emph{Roll-back attacks:}
As explained in Section~\ref{sec:implementation-rate_limiting}, our implementation can detect roll-back attacks on sealed data, using the platform's hardware-backed monotonic counters.
It can also detect and penalize any attempts to circumvent this protection (e.g.\ crashing the enclave).
While the enclave is running, all enclave data (even that paged out of the EPC) enjoys automatic roll-back protection provided by the CPU~\cite{Gueron2016}.

\emph{Run-time attacks:}
Remote attestation ensures that the \SafeKeeper enclave was started from trustworthy executable code.
However, if an adversary finds vulnerabilities in this code, he may attempt to mount run-time attacks.
The \SafeKeeper enclave itself contains only 248~lines of C code, excluding the Intel SDK.
We assume the SDK code is not vulnerable to run-time attacks, or at least has been thoroughly audited since it is commonly used by all SGX enclaves.
Therefore, \SafeKeeper's small size reduces the likelihood of vulnerabilities that could lead to runtime attacks, and makes \SafeKeeper amenable to security audits or even formal code verification.

\emph{Side-channel attacks:}
Various side-channel attacks have been demonstrated against SGX (e.g.~\cite{Xu2015,VanBulck2017}).
The \SafeKeeper enclave must protect four types of assets: 
(i) the plaintext user password; 
(ii) the \SafeKey; 
(iii) the channel keys resulting from the DH key agreement; and 
(iv) the DH private values.
When performing any AES-based operations (e.g.\ computing a CMAC value, or decrypting a password using the channel key), the \SafeKeeper enclave uses hardware AES-NI instructions, which are not vulnerable to software side-channel attacks\footnote{\url{https://software.intel.com/en-us/node/709048}} including the cache timing attack by G\"{o}tzfried et al.~\cite{Gotzfried2017} against a software AES implementation.
When performing the DH key agreement, \SafeKeeper uses Intel's trusted key exchange functions, which, to the best of our knowledge have not been shown to be vulnerable to side-channel attacks.

\subsection{Usability Evaluation}
\label{sec:evaluation-usability}

As part of evaluating \SafeKeeper against our Requirement~\ref{R2}, we conducted a user study for the client-side browser addon.
The primary objectives of this study include: (i) quantify the ability of participants to use the addon correctly; (ii) assess the \emph{memorability} of the addon usage after a period of disuse; and (iii) analyse the difficulty of using the addon. 
This user study was carried out in accordance with the standard practices of our institution.
No data protection issues arose because participants were not asked to use or disclose any sensitive personal data.

\subsubsection{Participants and methodology}
We recruited 86 participants using institutional mailing lists and social media. 
The participants were randomly split into two groups: main study group (64 participants), and control group (22 participants).
We collected basic demographic and background information for the main study group: 70\% of participants were male and 30\% were female; ages between 18 to 39 years; participants' educational qualification: 2\% Ph.D.\ 34\% Master's, 41\% Bachelor's, 9\% High school diploma (14\% unspecified).

\textbf{Main Group:} Each participant in this group was initially shown the \SafeKeeper information page, which contains the same information a normal user would see when installing our browser addon.
Participants were then given a set of 25 websites.
These websites were clones of popular websites, created using \emph{HTTrack}\footnote{\url{https://www.httrack.com/}} and hosted on an internal institutional server.
We slightly modified these sites for our experiment; see Table~\ref{table:websites}.
Out of the 25 websites, 9 used \SafeKeeper to protect the user's password, but 5 of these also spoofed the \SafeKeeper UI on some other input field in an attempt to confuse users.  Of the 16 websites that did not protect the password, 4 used \SafeKeeper to protect a non-password input field while showing the addon icon; 10 websites spoofed the UI to highlight the password field, and 3 of these spoofed the UI after a time delay (i.e.\ the attack explained in Section~\ref{sec:implementation-scripts}).
The order in which the websites were listed was the same for all participants, but participants could access these in any sequence and had the option to return to previous websites.
We did not inform participants anything about spoofing.

\begin{table}[ht]
\centering
\caption{Test websites}
\label{table:websites}
\resizebox{\columnwidth}{!}{\begin{tabular}{|p{16mm}|p{14mm}|p{38mm}|c|}
\hline 
\textbf{SafeKeeper lock icon} & \textbf{Password protected} &
\multicolumn{1}{c|}{\textbf{Spoofing types}} & \textbf{\# tests} \\ \hline
\ccmarkl & \ccmarkm                         & None                                           & 4                           \\ \hline
\ccmarkl & \ccmarkm                         & Other input fields highlighted                 & 5                           \\ \hline
\cxmarkl & \cxmarkm                          & None                                           & 6                           \\ \hline
\cxmarkl & \cxmarkm                          & Password field highlighted                     & 3                           \\ \hline
\ccmarkl & \cxmarkm                          & Password field highlighted                     & 4                           \\ \hline
\cxmarkl & \cxmarkm                          & Password field highlighted after delay         & 3                           \\ \hline
\end{tabular}}
\end{table}

For each website, participants were asked: \emph{``Does this website protect your password using \SafeKeeper?''}
Participants were instructed not to enter any password or other information on the website, but simply to record their answer on the provided paper form. 
The available options were: \emph{Yes}, \emph{No}, and the level of certainty of the answer, ranging from 1 (least certain) to 4 (most certain).
In order to assess the statistical significance of our results, we assumed a null hypothesis in which participants guess either \emph{Yes} or \emph{No} in a uniformly random manner.
Under this null hypothesis, the effectiveness would therefore be 50\%.
We assumed the standard 5\% threshold value for statistical significance ($\alpha = 0.05$).

\textbf{Follow-up Study:} After two months, we invited 20 participants with the highest scores (in terms of successfully identifying \SafeKeeper protection) in the initial study to participate in a follow-up study.
The objective of this study was to measure how well the participants remembered how to use \SafeKeeper, after a relatively long period of disuse.
The procedure was the same as for the initial study, except that participants were not shown the original \SafeKeeper information page, and were not reminded
of the instructions for using \SafeKeeper.
They were asked to visit the same set of websites in a different order and to answer the same question.

\textbf{Control Group:} To obtain a baseline against which to assess the results of the follow-up study, we invited approximately the same number of participants
(22) to use the tool without any instructions.
This was the same procedure as the follow-up study, except that this control group had never previously used the addon.
This control group is therefore the best approximation of the scenario in which users have completely forgotten how to use \SafeKeeper.

\subsubsection{Results}

In the main group, participants correctly identified 86.81\% of websites.
The average level of certainty per website varied from 3.5 to 4.
Calculating the $p$-value for each participant (across all websites) showed that for 80\% of participants, $p < 0.001$.
The $p$-value for each website (across all participants) does not exceed $0.001$ for 88\% of websites.

In the initial study, the top 20 participants correctly identified 93\% of websites on average.
In the follow-up study, they correctly identified 91\% of websites on average.
Calculating the $p$-value for each participant using McNemar's test shows that for 95\% of participants, the $p$-value exceeds 0.2.
This means that there was no statistically significant decrease in the effectiveness of \SafeKeeper after two months of disuse.
The $p$-value of only one participant shows a statistically significant decrease in effectiveness.

Even with no instructions, the control group participants correctly identified 74\% of websites on average.
Using Fisher's exact test to compare the follow-up group to the control group resulted in $p=6.4\times10^{-14}$ (far smaller than the threshold of 0.05), showing that the difference is statistically significant.

\subsubsection{Discussion}

Achieving almost 87\% effectiveness, we have exceeded the percentage indicated in the null hypothesis, providing evidence of the addon's utility.
Out of 64 participants in the main group, only for 5, $p > 0.05$.
Among the 25 websites, only for one website, $p > 0.05$
(the first phishing website participants encountered).

The follow-up study indicates that the effectiveness with which users use \SafeKeeper may not diminish, even after long periods of disuse (only one out of 20 participants showed a statistically significant drop in effectiveness).

Surprisingly, the control group managed to use the browser addon without any instructions to correctly identify a relatively high number of websites.
We suspect that these participants may have inferred the instructions to a certain extent based on the addon's behaviour and the text displayed in the popup window (Figure~\ref{fig:popups}). 

Based on the data gathered from the background questionnaires, nearly 58\% of participants reported that they do not usually check for a secure connection while browsing the web.  
75\% of participants were aware of phishing.
When asked to assess the level of difficulty of using \SafeKeeper, participants answers were: ``very easy to use'' (39\%), ``easy to use'' (55\%), ``difficult to use'' (6\%), ``very difficulty to use'' (0\%).  
Overall, 76\% of participants said they would like to use \SafeKeeper in their own browsers.
However, five participants mentioned that they would not use \SafeKeeper since they are ``not concerned about the security of passwords.''
Nine participants indicated that they would like to know more about the underlying technology.

\subsection{Performance Evaluation}
\label{sec:evaluation-performance}

Using the implementation described in Section~\ref{sec:implementation}, we evaluated the \emph{memory consumption} and \emph{scalability} of our server-side password protection service, as well as the \emph{latency} of verifying a quote with the IAS via our attestation proxy.
All reported performance measurements are the average of 10~trials, using a simulated password database of 1~million active unique users (i.e.\ 1~million unique salt values), performed on an Intel Core i5-6500 3.20~GHz CPU with 8GB of RAM.

\emph{Memory consumption:} Our enclave required at most 110~MB of heap memory to store an in-memory rate-limiting map containing all 1~million salt values.
Although this may exceed the available EPC size on some platforms, SGX supports paging of EPC pages, so the number of users that can be supported by a single platform is limited only by the platform's total available memory.
As discussed in Section~\ref{sec:implementation-rate_limiting}, memory consumption could be reduced by using more compact representations of the salt or counter values.
Note that the enclave does not have to store the salts of every user in the database at the same time -- it only stores salts it has processed in the current time window (e.g.\ users who authenticated in the past 24~hours).
The memory consumption can thus be further decreased by reducing this time window (e.g.\ setting the rate to 36 attempts per 6~hours would only require storing salts for the past 6~hours).

\emph{Scalability:} To measure the performance of the server when multiple users attempt to authenticate concurrently, we instrumented \texttt{PHPass} to measure the maximum password processing rate.
With the default hash function (software MD5, 256 iterations), the maximum rate is 446 ($\pm$10) passwords/second.
With the \SafeKeeper password protection service, the rate \emph{increases} to 1653 ($\pm$70) passwords/second, since we do not require multiple iterations.
We also measured the raw performance of the enclave without PHP (e.g.\ for websites running optimized software): the maximum rate is 101,337 ($\pm$4186) passwords/second.
Therefore, even for high-volume websites, this is unlikely to be a performance bottleneck.

\emph{Latency:} The average latency of verifying a quote with the IAS via our attestation proxy is 866~ms ($\pm$25~ms), average over 10 trials). 
If the connection to the proxy is re-used (e.g.\ when the user opens a new tab in the same browser window), this decreases to 296~ms ($\pm$5~ms).
This quote verification can be done asynchronously by the \SafeKeeper browser addon while the web page loads, and the latency is comparable with the typical loading time of the web page, this may not have any noticeable impact on the user's browsing experience.

\subsection{Deployability Evaluation}
\label{sec:evaluation-deployability}

\emph{Minimal software changes:}
\SafeKeeper's server-side password protection service is designed to be a drop-in replacement for the one-way functions used in current password hashing frameworks (e.g.\ the \texttt{PHPass} library).
Integrating the password protection service into the \texttt{PHPass} library required adding one line of code to initialize the enclave, and changing three lines of code in the password processing function.

Existing servers using the \texttt{PHPass} library for password hashing can be gradually migrated to \SafeKeeper. 
First, the necessary hardware support should be added to provide the SGX TEE. 
Next, the \texttt{PHPass} library is updated to the \SafeKeeper version.
Then all password hashes in the existing database are input to the \SafeKeeper enclave, which produces new hash values that are stored in the database.
This results in a so-called \emph{onion hash} in which the password is first hashed by the original \texttt{PHPass} library function, and then the result is processed by the \SafeKeeper password protection service.
This approach allows the server upgrade to be completed without requiring users to log-in or provide their passwords.
Finally, the web pages should be annotated to specify which fields are protected by \SafeKeeper.
This process is transparent to end users and requires only small changes on the server side.

\section{Extensions and Variations}
\label{sec:extensions}

In this section, we propose several extensions and variations of \SafeKeeper for other use-cases.

\subsection{Backup and Recovery}
\label{sec:extensions-backup}

In real-world deployments, server hardware sometimes fails.
Without an adequate backup and recovery procedure, failure of SGX hardware would result in loss of the \SafeKey, forcing all users to change their passwords.
A simple backup protocol would be to copy the \SafeKey to a type of \emph{backup enclave} running on a different physical machine.
Using remote attestation, the primary \SafeKeeper enclave could ensure that the \SafeKey was entrusted to a legitimate backup enclave.
If the primary enclave fails, the backup enclave could restore the \SafeKey to a new \SafeKeeper enclave, after attesting the new enclave.
However, this naive protocol could be abused to subvert the online guessing rate-limiting mechanism: the adversary could restore the \SafeKey to another machine, or even to multiple machines, even if the primary had not failed, thus increasing his effective guessing rate.

To prevent this attack, \SafeKeeper uses a backup and recovery protocol based on \emph{unanimous approval} of all key-holding enclaves.
Initially, the only key-holding enclave is the primary enclave $p_0$, since it generated the \SafeKey.
To create a backup, a \SafeKeeper backup enclave $b_0$ is launched on a separate physical machine, and attested by $p_0$ (e.g.\ the \texttt{MRENCLAVE} value of valid backup enclaves could be hard-coded into $p_0$).
If the attestation succeeds, $p_0$ establishes a secure channel to $b_0$ and sends it the \SafeKey, and a list of all current key-holding enclaves $\{p_0,b_0\}$, signed by $p_0$.
A correct backup enclave $b_i$ will only accept the \SafeKey if it is accompanied by matching signed lists from all key-holding enclaves, and all lists include $b_i$.
Similarly, a new primary enclave $p_i$ will only accept a \SafeKey if it is accompanied by matching signed lists from all key-holding enclaves that all designate $p_i$ as the primary enclave.
Since the enclaves themselves are guaranteed to behave correctly, this maintains the invariant that \emph{all key-holding enclaves are aware of all other key-holding enclaves}.
If an enclave does not receive all the required signed lists, it will not do anything.

However, if a machine fails, its enclave can no longer provide an updated list in order to change the set of key-holding enclaves.
In this case, the operator can provide all other key-holding enclaves with a proof that the failed machine has been revoked (e.g.\ a signed assertion from Intel).
All key-holding enclaves can safely remove the failed machine (either a primary or backup) from their lists.
For example, if $p_0$ fails and is revoked, $b_0$ can update its list to $\{b_0\}$.
A new primary enclave $p_1$ can be created and attested by $b_0$, which then sends $p_1$ the \SafeKey and an updated list $\{p_1,b_0\}$ signed by $b_0$, thus allowing $p_1$ to start operating.

In terms of security, the adversary can always prevent one or more enclaves from providing such lists, but this only leads to denial of service, which does not threaten the security of the passwords.
An enclave that has not actually failed can still be revoked by a malicious operator (e.g.\ in order to activate a new primary concurrently).
To prevent this, each enclave checks whether it has been revoked (e.g.\ whenever the \texttt{reset\_attempts()} function is called).

This protocol also allows multiple backups to be created $b_0, ... b_n$, thus improving redundancy whilst still maintaining the invariant.
In the example above, if $p_1$ wants to make a second backup $b_1$, the process is the same except that $b_1$ must receive two lists of the form $\{p_1,b_0,b_1\}$, signed by $p_1$ and $b_0$.
Although users only attest the primary enclave, they can \emph{transitively} trust the backup enclaves, because these have been attested by the primary.
The code used to attest the backup enclaves is included in the quote of the primary, which is checked by the users.

\subsection{Scaling to Multiple Servers}
\label{sec:extensions-scaling}

Although our performance evaluation shows that \SafeKeeper is unlikely to be a performance bottleneck even for high-volume websites, a website operator may still wish to run \SafeKeeper's password protection service on multiple physical machines (e.g.\ in different geographic locations to reduce latency).
Similarly to the backup and recovery protocol, simply duplicating the primary enclave on another machine would increase the adversary's rate of guessing, as he could guess on both primaries.

To support \emph{secure} scaling to multiple servers, the backup and recovery protocol described above can be extended to support \emph{multiple concurrent primaries}.
The overall permissible password checking rate is divided among the primaries.
The rate of each primary is included in the list of key-holding enclaves, and all enclaves must check that the sum of these rates equals the predefined total rate (i.e.\ 144 guesses per 24 hours in our implementation).
For example, with two concurrent primaries $p_0$ and $p_1$, and one backup enclave $b_0$, the list of all key-holders would be $\{p_0=72,p_1=72,b_0\}$.
If primary $p_0$ now fails, $p_1$ and $b_0$ could remove it from their lists after receiving the signed revocation statement.
The rate of $p_1$ could then be increased if $b_0$ signs the list $\{p_1=144,b_0\}$.

As in the backup and recovery protocol, \emph{increasing} the rate of any primary requires unanimous approval of all key-holding enclaves.
However, \emph{decreasing} the rate of any enclave (even down to zero) does not require any approval (i.e.\ an enclave decreases its own rate before signing a new list that would cause its rate to decrease).
If the decrease operation required unanimous approval, the adversary might simply block some of the messages in order to maintain the higher rate.
In this way, the overall rate still cannot increase beyond the maximum, but also does not decrease in the event of hardware failures.

To simplify the implementation and the attestation of enclaves, there is no difference between a primary enclave and a backup enclave in our design -- backup enclaves are just primary enclaves with a rate of zero.
This means that even in the simplest scenario of one primary and one backup, the backup can automatically become a primary and start processing requests as soon as the original primary fails.

\subsection{Protecting Email Address Confidentiality}
\label{sec:extensions-email}

Many websites use email as a password reset mechanism e.g.\ a password reset token can be sent to the user's registered email address.
However, this requires the server to store users' email addresses, which could lead to abuse by a malicious server operator (e.g.\ spamming the user) or the email addresses could be leaked through a data breach.
If the email address is only needed for password recovery, a variant of \SafeKeeper can protect it against unauthorized disclosure, even by a malicious server operator or an adversary who compromises the server.
This variant may also be useful for other similar privacy-sensitive information.

\begin{figure}[ht]
	\begin{center}
    \includegraphics[width=1\columnwidth]{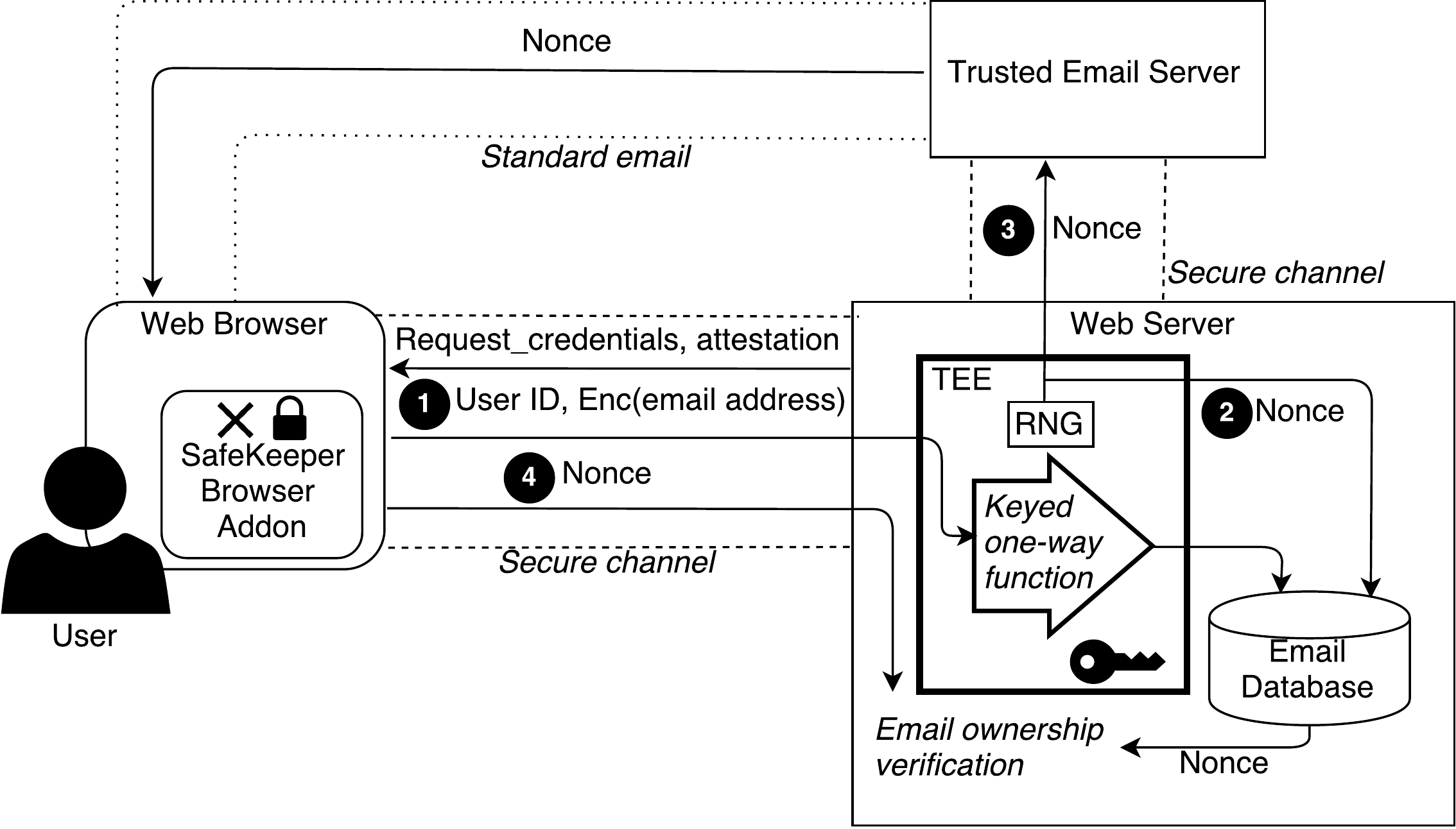}
    \caption{Overview of the design of the email protection service.}
    \label{fig:email}
	\end{center}
\end{figure}

As shown in Figure~\ref{fig:email}, this makes use of a new server-side component, the \SafeKeeper \emph{email protection service} (these TAs can be merged if needed).
Like the password protection service, this component is also isolated using a TEE.
The email protection service is used during user registration, and password reset, however, the behaviour of the service is the same in both cases.
When the user registers, her email address is encrypted by \SafeKeeper's client-side software in the same way as a password, and is sent to the email protection TA \dOne.
The TA decrypts and then computes a keyed one-way function of the email address, and outputs the result to the server, which stores it in a database.
If email ownership verification is required (indicated by the web server), the email service generates a random nonce corresponding to the email address, and outputs this to the server \dTwo.
The email protection service then establishes a secure connection to a trustworthy third party email provider (e.g.\ ProtonMail\footnote{\url{https://protonmail.com/}}), and sends the nonce to the user via email \dThree.
The TA establishes a secure connection to the third party email provider (e.g.\ using the TaLoS TLS library~\cite{Aublin2017} or the Intel SGX SSL crypto library\footnote{\url{https://github.com/01org/intel-sgx-ssl}}) so that the email address is not revealed to the server.
Upon receiving this email, the user visits the registration web page hosted on the server and provides the random nonce she received via email to prove ownership \dFour.  
To improve usability, the nonce could be encoded into the verification URL (e.g.\ like an existing password reset link).

When a user forgets her password, the process is similar, and the functionality of the email protection service is unchanged.
The user sends her email address to the email protection service, which computes the same keyed one-way function and outputs the result to the server.
The server checks whether that particular result is already in its database (i.e.\ if the user has actually registered). 
If not, the server returns an error to the user, but if the address is valid, the untrusted server asks the email protection service to proceed with ownership verification.
If verification is successful, the user is allowed to reset her password (via \SafeKeeper).
This works for web sites that require the user to re-enter her email address in the password reset process.
For web sites that allow password reset based only on user ID, an alternative design would be for \SafeKeeper to seal (i.e.\ encrypt) the email addresses during registration and then unseal them when a password reset is required.
Using this extension does not preclude websites from using the email address as the user ID -- from the website's perspective, the user ID simply becomes the result of the one-way function applied to the email address.

\emph{Note:} This process does not prevent a rogue server from changing users' passwords without the users' consent -- this is already possible because the adversary can execute arbitrary code on the server.
Our goal is only to protect the confidentiality of users' passwords and email addresses.

This type of construction could also be used to protect other types of sensitive credentials.
For example, users' phone numbers could be protected in the same way by establishing a secure connection between the \SafeKeeper enclave and a trustworthy voice-over-IP (VoIP) provider, thus allowing the enclave to e.g.\ initiate a call to users and play a pre-recorded audio message.
Similarly, payment card details could be protected by establishing a secure connection from the \SafeKeeper enclave to payment card providers.
However, the full design of these solutions is out of scope for this paper.

\subsection{Another Defence Against Malicious Scripts}
\label{sec:extensions-scripts}

\begin{figure}
	\begin{center}
    \includegraphics[width=0.3\columnwidth]{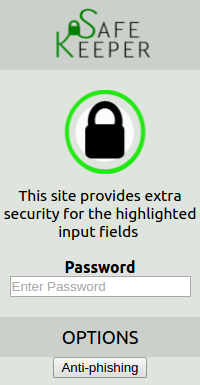}
    \caption{An alternative popup window with embedded password input field.}
    \label{fig:popup2}
	\end{center}
\end{figure}

Section~\ref{sec:implementation-scripts} outlined the threats posed by malicious client-side scripts, and the challenges faced in mitigating these (i.e.\ possible race conditions, and reduced web page functionality).
An alternative to disabling scripts is to take password input directly through a \SafeKeeper controlled popup window; see Figure~\ref{fig:popup2}.
Text entered into this field is always encrypted and sent directly to the \SafeKeeper TA; malicious scripts at the client cannot spoof input fields or read the typed passwords with this design.
The challenge is to teach users to enter passwords only into the \SafeKeeper window.
Also, this prevents legitimate client-side scripts from reading the password (e.g.\ evaluating password strength, or enforcing password complexity policies).
To maximize usability and deployability, this variant should be optional (e.g.\ activated on-demand by a button in the popup window).

\subsection{No Click Required}
\label{sec:extensions-click}
We currently require users to check for \SafeKeeper protected fields by clicking on an icon (recall Figure~\ref{fig:icons}).
To alleviate this restriction, we propose another variant to show the \SafeKeeper lock symbol only when \emph{all} fields of a given page are protected. 
This may require some changes to the website so that users are shown a dedicated password entry page containing only the password input field.
However, some major websites already use this pattern (e.g.\ the Google sign-in page).
Alternatively, if there are multiple input fields on the page, it may be possible to route all data through the enclave and have the enclave release only non-sensitive data to the server.
However, this would require an automated method for the enclave to distinguish between passwords and non-sensitive data, which is beyond the scope of our current work.
In any case, the benefit of this approach is that the user only has to \emph{check} for the presence of the correct \SafeKeeper icon before entering a password, similar to the HTTPS lock icon. 
We envision that browser vendors may integrate \SafeKeeper availability (or TEE-based protection in general) into browser GUI, potentially within the existing HTTPS indicators.

\section{Related Work}
\label{sec:related_work}

Password research has a rich history. 
Here we focus on proposals that improve password security against password database compromises, emphasizing solutions that leverage hardware-based security enhancements such as Intel SGX.
However, to the best of our knowledge, no previous work has considered password confidentiality against rogue servers.

To mitigate offline guessing attacks against hashed passwords, e.g.\ due to password database breaches, several proposals take advantage of SGX enclaves~\cite{BirrPixton2016, Brekalo2016, Krawiecka2016}.
These proposals either encrypt or calculate an HMAC of the password inside an enclave in order to protect the secret key.
However, they all assume a significantly weaker adversary model in which the server operator is trusted.
Specifically, they do not consider the confidentiality of passwords on the server before they are input to the enclave, nor the possibility for a malicious server operator to perform a brute-force guessing attack using the enclave, nor the risk of phishing.
\SafeKeeper goes beyond these solutions by resisting rogue servers and phishing attacks.

Various non-SGX server-side approaches have been proposed.
Facebook uses a \emph{remote password service}~\cite{Everspaugh2015} that computes a keyed one-way function on the password and returns the result.
The secret key used in this function never leaves their service.
Even if an adversary obtains the (keyed) password repository, he must guess the secret key, or connect to the Facebook remote password service (easy to detect and rate-limit).
However, running such a service and protecting the key sever against attacks may be infeasible, especially with a limited budget.
Also, the key service must be trusted (e.g.\ not to collude with an attacker), but the trustworthiness of the key server cannot be validated by users.

Another server-side approach is to use specialized hardware to establish an isolated execution environment.
Cvrcek et al.~\cite{Cvrcek2014} use a custom-built USB device to store a secret key and use this to calculate HMACs of passwords.
Without the USB device, calculating the HMACs is not possible, and thus cracking passwords from leaked tags is also infeasible.
Their prototype was capable of computing only 330 HMAC tags per minute, although scalability could be improved using multiple USB devices. 
However, plaintext passwords remain available to the server operator, and the trustworthiness of the USB device cannot be validated by users.

As an alternative software-only server-side approach, PolyPasswordHasher~\cite{Torres2014} uses a special set of \emph{protector} account passwords to prevent offline dictionary attacks against the rest of the \emph{shielded} account passwords.
For protected accounts, it applies an XOR function to a \emph{share} and the hash of a salted password.
The share values are derived using a threshold cryptosystem, and stored only in memory (in plaintext).
The result of the XOR operation is stored in a database on the web server.
To guess shielded passwords, the adversary must crack a threshold of protector passwords (e.g.\ 3--5), and collect the corresponding shares.
The shielded passwords are encrypted using a secret, not the shares, and then stored in the database.
Therefore, the shielded accounts that use
weaker passwords will not leak the information about the shares.
The security of this solution highly depends on the passwords chosen by the protectors (e.g.\ admin accounts).

Users often remain logged into multiple services at the same time. This fact is leveraged in SAuth~\cite{Kontaxis2013} to detect password
database compromise.
A login attempt to a target site must be validated by a separate \emph{vouching} site (e.g.\ to log into Gmail, the user must be logged into Facebook).
Thus, an attacker who compromises passwords from one site, will be unable to use them, unless he can also compromise passwords from the vouching site (assumed to be unlikely).
 
Shirvanian et al.~\cite{Shirvanian2017} designed SPHINX to restrict attacks on password managers.
It is based on a new cryptographic primitive called device-enhanced password-authenticated key exchange (DE-PAKE~\cite{Jarecki2016}), and uses a unique randomized password for each site; the user memorizes only a master password.
The master password and site passwords remain safe under a compromised/malicious server-operator or manager component (implemented on a smartphone or as an online service), but not under compromise of both.
If a site password and the manager are compromised, an offline attack on the master password is possible (and consequently, other site passwords can also be derived).
Unlike in \SafeKeeper, a plaintext site password remains available to the site operator.

Online password manager Dashlane has partnered with Intel to enable SGX protection to user passwords by locally sealing the user's password database~\cite{dashlane-sgx}.
This will prevent guessing attacks even if the database is compromised from Dashlane.
As of writing (August 2017), SGX support is yet to be implemented, and it is unclear how such protected password be accessed from other devices.
Intel also provides an independent open-source local password manager.\footnote{\url{https://github.com/IntelSoftware/Tutorial-Password-Manager-with-Intel-SGX}}

Password database compromise in the case of password managers (especially for online managers) is far more devastating than a single site compromise.
Several solutions have been proposed recently, including attacks against them; see e.g.~\cite{Bojinov2010, Juels2013, Chatterjee2015, Golla2016}.
In general, these solutions rely on decoy passwords (generated following realistic password patterns).
After a trial decryption of a password vault, an attacker cannot readily distinguish if the candidate password list is real or decoy, unless he contacts an online service (which can be detected). 
Attacks against these solutions (e.g.\ \cite{Pasquini2017}) generally exploit differences in real and decoy passwords, leveraging limitations in decoy password generation.

Several client-side techniques have also been proposed.
As an example, the PwdHash~\cite{Ross2005} browser addon applies a pseudorandom function to the concatenation of a password and a salt, on the client side.
The salt is generated based on a website domain, which helps to bind the password to the specific website.
However, an adversary who compromises the web server knows the domain name, and can therefore include this when mounting a brute-force guessing attack.

Compared to existing work, \SafeKeeper provides additional security features, including: protecting password confidentiality against a malicious server operator, strict password guess rate-limiting from within our enclave, enabling users to validate SGX protections of their passwords via browser UI, and relatively easy deployment options for server operators.

\section{Conclusion and Future Work}
\label{sec:conclusion}

Passwords are likely to remain the de facto approach for authenticating users on the web, despite their inherent security weaknesses.
Therefore, it is critical to improve the security of such systems without decreasing performance, usability, or deployability.
A comprehensive solution must address the dual threats of phishing and theft of password databases, even in the case of rogue servers. 
We have demonstrated that \SafeKeeper is a significant step towards meeting these objectives.

As future work, we plan to implement selected extensions and variations, as described in Section~\ref{sec:extensions}.
We also plan to integrate \SafeKeeper's client-side assurance mechanism into other types of client-side software, beyond the web browser.
For example, \SafeKeeper could be integrated with client-side password managers, such as Dashlane~\cite{dashlane-sgx}.
The password manager itself could attest the server-side password protection service, and agree on a session key, without relying on the browser addon.
In this way, the password manager could encrypt the password with the session key before it is entered into the web page, thus preventing the password from being snooped by malicious client-side scripts.
Furthermore, if the password manager uses a TEE on the client (e.g.\ as planned in Dashlane~\cite{dashlane-sgx}), the key agreement and encryption could be done inside the client's TEE.
This would protect the confidentiality of the user's passwords against rogue servers \emph{and} malicious client-side software.

\begin{acks}
This work was supported in part by the Intel Collaborative Research Institute for Secure Computing at Aalto University, and the Cloud Security Services (CloSer) project (3881/31/2016), funded by Tekes/Business Finland.
M.~Mannan is supported in part by an NSERC Discovery Grant and a NordSecMob Scholarship.
The authors thank Paul van Oorschot and Michael Steiner for helpful discussions about this work.  
\end{acks}

\balance

\raggedright
\bibliographystyle{ACM-Reference-Format}
\bibliography{sgx-passwords}

\end{document}